\documentclass[12pt,a4paper]{article}
\usepackage{amssymb,amsfonts,amsmath}
\usepackage{url}

\begin{document}

\title{Relational Quantum Mechanics:  Ozawa's Intersubjectivity Theorem  as justification of the postulate on internally consistent descriptions}

\author{Andrei Khrennikov\\ 
Linnaeus University, International Center for Mathematical Modeling\\  in Physics and Cognitive Sciences
 V\"axj\"o, SE-351 95, Sweden}

\maketitle

\abstract{The Ozawa's Intersubjectivity Theorem (OIT) proved within quantum measurement theory supports the new postulate of  relational quantum mechanics (RQM), the postulate on internally consistent descriptions. We remark that this postulate was proposed only recently to resolve the problem of intersubjectivity of information in RQM. In contrast to RQM for which OIT is a supporting theoretical statement, QBism is challenged by OIT.}     

{\bf keywords:}  Relational Quantum Mechanics, Ozawa Intersubjectivity Theorem, postulate on internally consistent descriptions, 
 measurement process, probability reproducibility, measurement in the same basis

\section{Introduction}

Recently Ozawa's Intersubjectivity Theorem (OIT) \cite{OIT} proved within quantum measurement theory (e.g., \cite{VN}-\cite{O4}) was used as objection to QBism \cite{Qozawa}, to its basic interpretational statement about the individual agent perspective on measurement's outcomes \cite{Fuchs1}-\cite{Fuchs3}. (This privacy component of QBism was also criticized in my previous articles \cite{KHRQBism1,KHRQBism2}, mainly from the probabilistic perspective.) Now QIT provides the strong argument that measurement's outcome is intersubejctive and, although one can still  interpret it as a private experience, such interpretation loses its elegance. QBists replied to my objection with two papers \cite{{QBismReply1},QBismReply2} (the second one presents more elaborated counterargument).

In a few words the QBism's counterargument is that compatible
 measurement devices $M_1$ and $M_2$ considered in OIT should be related not to  two distinct observers $Q_1$ and $O_2$ as was done 
in my article \cite{Qozawa}, but to one observer $O$ who performs these two measurements.  As was clearly explained in papers \cite{QBismReply1,QBismReply2}, this interpretation of such measurements is consistent with the general ideology of QBism. The important impact of article \cite{Qozawa} to QBism is that this article attracted the attention (at least some) of QBists to 
quantum measurement theory \cite{VN}-\cite{O4}. QBism intensively explores one of the basic elements of this theory - POVMs. 
But QBists don't employ the important part of measurement theory - theory of {\it quantum measurement processes.} Each POVM is generated by 
a measurement process \cite{O1}. It is important to note that the same POVM can be generated by variety  of measurement processes.  

The discussions on the intersubjectivity problem in QBism stimulated my interest to this problem for other information interpretations of QM. A reviewer of one of my papers pointed that relational quantum mechanics (RQM) \cite{RQM0}- \cite{RQMI} was suffering of the same problem. It seems that it can't be resolved within ``old RQM''. Its resolution requires invention of an additional postulate on 
{\it internally consistent descriptions} (Postulate 6 in \cite{RQMI}). By treating this postulate within quantum measurement theory, one immediately see that OIT supports this postulate. So, in contrast to QBism which was challenged by OIT,   RQM's  Postulate 6  is mathematically justified via OIT. However, clarification of the interrelation of the conditions ``probability reproducibility'' and 
``measurement in the same basis''  is needed (see section \ref{P6} and section \ref{PR} for more detail).

As well as QBism,  RQM can earn from closer connection with quantum measurement theory \cite{VN}-\cite{O4}. Besides the present paper,  an attempt to proceed in this direction was done in article \cite{Pekka3}. However, the conclusions of this article are debatable. I am not an expert in  RQM and I am not able to evaluate whether the authors of 
\cite{Pekka3} correctly understood the conceptual premises of RQM.  

We start with a brief recollection of the basics of RQM and the problem of intersubjectivity (section \ref{RQ}). We follow article \cite{RQMI}. Then we formulate OIT (section \ref{P6}), see article \cite{OIT} for its proof. In section \ref{P6} we 
couple OIT with Postulate 6 (on intersubjectivity) of RQM.  Matching of quantum measurement theory and RQM is established in section \ref{apparatus}. We should modify the interpretation  and representation of the original von Neumann's measurement scheme to make it 
consistent with the RQM perspective on observer: \begin{footnotesize} 
``Any system, irrespectively of its size, complexity or else, can play the role of the textbook's quantum mechanical observer.''   
\end{footnotesize} (See \cite{Stanford}).\footnote{This is the good place to cite another block from article  \cite{Stanford}:
``In textbook presentations, quantum mechanics is about measurement outcomes performed when an “observer” makes a “measurement” on a quantum system. What is an observer, if all physical systems are quantum? What counts as a measurement? Common answers invoke the observer being macroscopic, onset of decoherence, irreversibility, registration of information, or similar. RQM does not utilise anything of the sort.''}  

\section{Relational quantum mechanics and the problem of intersubjectivity}
\label{RQ}

As was stated by Rovelli \cite{RQM0}, RQM is based on the idea that 
 \begin{footnotesize}
``in quantum mechanics different observers may give different accounts of the same sequence of events.''\end{footnotesize}
It is crucial that 
 \begin{footnotesize} ``RQM is built on strong naturalistic intuitions; therefore, in RQM, the term `observer' is
understood in a broad sense, which allows that any physical system can be an `observer,' so we do not have to accept that consciousness plays any fundamental role.'' \end{footnotesize}\cite{RQMI}. In the last aspect RQM differs crucially from QBism. 
As is emphasized in article \cite{RQMI}, RQM has many attractive features and it resolves the basic quantum paradoxes (but QBism 
do this as well).   

\begin{footnotesize}
``However, some problems remain; in particular, there is a tension between RQM’s
naturalistic emphasis on the physicality of information and the inaccessibility of certain
sorts of information in current formulations of RQM. Thus, in this article, we propose
a new postulate for RQM which ensures that all of the information possessed by a
certain observer is stored in physical variables of that observer and thus is accessible by
measurement to other observers. The postulate of cross-perspective links makes it
possible for observers to reach intersubjective agreement about quantum events that have
occurred in the past, thus shoring up the status of RQM as a form of scientific realism and
allowing that empirical confirmation is possible in RQM.

Adding this postulate requires us to update some features of the ontology of RQM, because
it entails that not everything in RQM is relational.''\end{footnotesize}\cite{RQMI}

For readers convenience, here the list of all RQM's postulates is presented \cite{RQMI}:

\begin{enumerate}
\item Relative facts: Events, or facts, can happen relative to any physical system.
\item No hidden variables: Unitary quantum mechanics is complete.
\item  Relations are intrinsic: The relation between any two systems $A$ and $B$ is
independent of anything that happens outside these systems' perspectives.
\item Relativity of comparisons: It is meaningless to compare the accounts relative to
any two systems except by invoking a third system relative to which the comparison
is made.
\item   Measurement: An interaction between two systems results in a correlation within
the interactions between these two systems and a third one; that is, with respect to
a third system $W,$ the interaction between the two systems $S$ and $F$ is described by a
unitary evolution that potentially entangles the quantum states of $S$ and $F.$
\item Internally consistent descriptions: In a scenario where $O_1$ measures $S,$ and $O_2$ also
measures $S$ in the same basis, and $O_2$ then interacts with $O_1$ to ``check the reading'' of
a pointer variable (i.e., by measuring $O_1$ in the appropriate ``pointer basis''), the two
values found are in agreement.
\end{enumerate}

OIT matches with Postulate 6 of RQM and justifies it mathematically on the basis of quantum measurement theory.  Generally  RQM can earn a lot from employing  quantum measurement theory.

As was cited above in  \begin{footnotesize}``RQM the term `observer' is understood in a broad sense, which allows that any physical system can be an `observer' ...''\end{footnotesize}

In quantum measurement theory, `apparatus' plays the key role. I would suggest to extend its meaning and operate in this theory with the notion observer. The latter has the RQM meaning, i.e., any physical system can be an observer. 
This the important novelty in application of quantum measurement theory and it will be discussed in more details (see section \ref{apparatus}).

\section{Postulate on internally consistent description in the view of Ozawa Intersubjectivity Theorem}
\label{P6}

Von Neumann \cite{VN} described observables mathematically by Hermitian operators acting in  Hilbert state space ${\cal H}.$
They represent {\it accurate measurements.}  Consider operators with totally discrete spectra $X \subset 
\mathbb R:  A=\sum_{x \in X} x E_A(x),$
where $E_A(x)$ is projection on the eigensubspace for the eigenvalue $x.$  
The Born rule determines the probabilities:
$$
P(A = x| \psi) = \langle \psi | E_A(x) |\psi \rangle.
$$

The indirect measurement scheme involves the following components
\begin{itemize}
\item the states spaces ${\cal H}$ and ${\cal K}$ of the systems ${\cal S}$ and the apparatus ${\cal M}$ for measurement of some observable $A;$     
\item the evolution operator $U=U(t)$ representing the interaction-dynamics for the system ${\cal S}+ {\cal M};$ 
\item the meter observable $M$  giving outputs of the pointer of the apparatus ${\cal M}.$ 
\end{itemize}

 Let $E_M=(E_M(x))$ be the spectral family of the operator $M;$ here $E_M(x)$ are  projections in  Hilbert space ${\cal K}.$  It is assumed that the compound system's  evolution is driven by the Schr\"odinger equation, so the  evolution operator $U(t)$ is unitary. 

Formally, an {\em indirect measurement model} for an observable $A$, introduced in \cite{O1} as a ``measurement process'', 
is a quadruple 
\begin{equation}
\label{MA00}
({\cal K}, |\xi\rangle , U, M),
\end{equation}
where  $|\xi\rangle \in {\cal K}$ represents the apparatus state.  

We explore the Heisenberg picture. To describe meter's evolution, we represent it in the state space of the compound system, i.e., as
$I \otimes M.$  The meter observable evolves as
\begin{equation}
\label{MA1}
M(t) = U^\star(t)(I\otimes M) U(t).
\end{equation} 
By the Born rule 
\begin{equation}
\label{MA2}
P(M(t)= x| \psi \xi) = \langle \psi \xi | E_{M(t)}(x)| \psi \xi \rangle .
\end{equation} 
This is the probability distribution for the outputs of measurements done by the apparatus and given by the meter.

{\bf Definition.} {\it A measurement process $({\cal K}, |\xi\rangle , U, M)$ reproduces the probability distribution for  quantum observable $A$  (accurate von Neumann observable)  if} 
\begin{equation}
\label{MA7a}
P(A = x| \psi) = P(M(T)= x| \psi \xi).
\end{equation} 

Following \cite{OIT}, consider two remote observers $O_1$ and $O_2$ who perform joint measurements on a system ${\cal S},$  in mathematical terms it means that the meter quantum observables of the corresponding measurement processes commute, 
 $$
[M_1(t), M_2(t)]=0.
$$ 
Here each apparatus has its own state space, i.e., 
${\cal K}= {\cal K}_1 \otimes {\cal K}_2.$ We call such measurements local. 
In this situation the joint probability distribution is well defined 
\begin{equation}
\label{MA9}
P(M_1(t)= x, M_2(t)= y | \psi \xi_1 \xi_2) = \langle \psi \xi_1 \xi_2| E_{M_1(t)}(x) E_{M_2(t)}(y)| \psi \xi_1 \xi_2 \rangle.
\end{equation}

\medskip

{\bf Theorem.} (OIT \cite{OIT}) {\it Two observers performing the joint local and probability reproducible  measurements of the same quantum observable $A$  on the system ${\cal S}$ should get the same outcome with probability 1:}
\begin{equation}
\label{MA10}
 P(M_1(T)= x, M_2(T)= y | \psi \xi_1 \xi_2) = \delta (x-y) P (E= x | \psi) = \delta(x-y) \Vert E(x) \psi \Vert^2.
 \end{equation}

In OIT,  as in Postulate 6 [RQM], observer $O_1$ measures $S,$ and observer $O_2$ also
measures $S$ in the same basis. In our terms the later means that they measure the same observable $A.$ Then $O_2$ can interact 
with $O_1$ to ``check the reading'' of its pointer variable, the two values found are in agreement. So, OIT mathematically justifies 
Postulate 6 [RQM]. Moreover, due to OIT observer $O_2$ even need not to perform a measurement on $O_1.$ If observers are sure in validity of quantum theory, then they can be sure that they get the same outcome.

In the previous version of this preprint and in published article \cite{KHR_RQM}, I claimed that 

\begin{footnotesize} ``The important condition which is missed in Postulate 6 is the prob-
ability reproducibility condition, only under this conditions outcomes
of measurements performed by $O_1$ and $O_2$ coincide. Therefore, it is
natural to complete postulate 6 by this condition.''\end{footnotesize}

This claim generated a mini-debate with professor Ozawa. In his comment he pointed out that Postulate 6 
\begin{footnotesize} ``has already been completed by the assumption of probability reproducibility formulated in, since the assumption ``$O_1$ measures $S,$ and $O_2$ also measures $S$ in the same basis'' usually means that both $O_1$'s measurement and $O_2$'s measurement satisfy the probability reproducibility condition for the observables determined by ``the same basis''.''
\end{footnotesize}

We come back to this discussion in section \ref{PR} after deeper presentation of von Neumann's measurement theory in section \ref{apparatus} by considering  the interrelation of  the conditions ``probability reproducibility'' and 
``measurement in the same basis''.

\section{Matching  von Neumann's  theory of quantum measurement with Rovelli's interpretation of observer} 
\label{apparatus}

As was emphasized in introduction, in this paper the meaning of the notion  ``apparatus'' is extended, we operate with the notion
``observer''.   And in RQM any physical system can play the role of an observer. The following question naturally arises:

{\it Are there some constraints on the notion of  ``apparatus" as it is defined in quantum measurement theory which might prevent it from encompassing all physical systems?} 

This paper is based on the modern version of von Neumann's approach to the problem of quantum measurement.  Therefore it is natural to return to the original von Neumann's work \cite{VN} (chapter 6). Our aim is to show that, in spite some interpretational differences,  are von Neumann's and Rovelli's  viewpoints on the quantum measurement process generally close to each other. In fact, elements of such comparative 
analysis can be found in the original article of Rovelli \cite{RQM0}. Our discussion is more detailed.
Moreover, we bring into our analysis  a new theoretical construction - ``entanglement of observables'', 
$A_1A_2$-entanglement for a pair of compatible observables. This generalization of entanglement was presented in articles 
\cite{ENT,ENT1}. It matches both with von Neumann's description of a measurement process and RQM's Postulate 5, on measurement. 
The probability reproducibility condition of OIT can also be formulated in terms of  ``entanglement of observables'', in the 
special case of von Neumann's model.  

We start with some interpretational remarks on measurement theory presented in \cite{VN}  (see section 1, chapter 6).
In this theory both a quantum system and measurement apparatus are described within the 
quantum formalism. Thus, although the apparatus  is macroscopic, it is, nevertheless, described as a quantum system.
Moreover, the mathematical model is symmetric w.r.t. system-apparatus interchange. Thus, it is applicable to macroscopic systems,
(cf. with RQM). This macro-applicability was recently supported by quantum-like modeling, applications of the mathematical formalism of quantum measurement theory outside of physics, e.g.,  in cognition and decision making; see \cite{ENTROPY,OJMP}. We also remark that von Neumann's description of the measurement process
doesn't involve decoherence (again cf. with RQM).

At the same time von Neumann's interpretation differs from RQM, since he refers to subjective 
nature of an observer (cf. with QBism). However, in contrast to QBists he didn't emphasize the private character of observer's
measurement experience, he neither use the subjective interpretation of probability. Moreover, he diminished the value 
of subjective experience by referring to psycho-physical parallelism. At the same time he states \cite{VN} (p. 420): 
\begin{footnotesize}   
``Indeed experience only makes statements of this type: an observer has made a certain (subjective) observation; 
and never any like this: a physical quantity has a certain value''
\end{footnotesize}
This is the Bohr's viewpoint on quantum observables \cite{BR0,PL2}: their outcomes aren't  the objective properties of systems under measurements. They are generated during the complex process of interaction between  system $S$ and  measurement apparatus 
${\cal M}$ used to measure quantum observable  $A.$  At this point the positions of von Neumann and Bohr coincide. In \cite{VN} interaction between the system and apparatus is mathematically described by unitary operator $U,$ see below. 

Now we turn to the formalism developed in chapter 6 \cite{VN} . The measurement process is described in section 3.
There are considered three systems: ${\cal S},$ observed system, ${\cal M},$ a measurement apparatus, and 
${\cal O}$, observer.  It is important to remark \cite{VN} (p. 421) that  \begin{footnotesize} `` ... ${\cal O}$ 
itself reminds outside of calculations.'' \end{footnotesize} (See alo p. 439). Hence, all mathematical considerations are related to the 
pair of physical systems, $({\cal S},{\cal M}).$ And in  section 3, ``measurement section'', von Neumann refers 
to calculations done in section 2 (``operator entanglement section'').

In fact, von Neumann considered a kind of entanglement which can be called  {\it ``entanglement of observables''} (see \cite{ENT,ENT1}). Two compatible observables $A_1$ and $A_2$ can be called  entangled in the state $|\Phi\rangle,$ if their 
eigenvalues can be enumerated in such a way, $(a_{1k}, k=1,2,...)$ and  $(a_{2k}, k=1,2,...)$ that probability
\begin{equation}
\label{ENT1a}
P(A_{1}= a_{1k}, A_{2}= a_{2m}| \Phi))=0, k \not= m,
\end{equation}
or 
\begin{equation}
\label{ENT2}
\sum_k P(A_{1}= a_{1k}, A_{2}= a_{2k}| \Phi))= 1,
\end{equation}
or 
\begin{equation}
\label{ENT2at}
P(A_{1}= a_{1k}| \Phi))= P(A_{1}= a_{1k}, A_{2}= a_{2k}| \Phi),
\end{equation}
\begin{equation}
\label{ENT2at1}
P(A_{2}= a_{2k}| \Phi)= P(A_{1}= a_{1k}, A_{2}= a_{2k}| \Phi),
\end{equation}
or
\begin{equation}
\label{ENT2a}
P(A_{1}= a_{1k}| A_{2}= a_{2k}| \Phi)= 1, P(A_{2}= a_{2k}| A_{1}= a_{1k}| \Phi)=1,
\end{equation}
under the constraint $P(A_{1}= a_{1k}, A_{2}= a_{2k}| \Phi) \not=0.$ Here conditional probability w.r.t. state $|\Phi\rangle$ is defined 
by the Bayes formula, e.g., $P(A_{1}= a_{1k}| A_{2}= a_{2k}| \Phi)= P(A_{1}= a_{1k}, A_{2}= a_{2k}| \Phi)/P(A_{2}= a_{2k}| \Phi)$
(since the operators commute we can operate within the classical probability framework).  

From my viewpoint, the notion of such $A_1A_2$-entanglement matches better with RQM, than the standard notion of the entangled state (see Postulate 5 of RQM).  
We remark that a state $|\Phi\rangle$ can be $A_1A_2$-entangled, but not entangled in the ordinary sense. Moreover, it was shown that, for any state  $|\Phi\rangle,$
there exist  pairs of operators which are entangled in this state \cite{VN} (section 2). We note that von Neumann didn't use the 
term ``entanglement'' for observables. He wrote about establishing ``one-to-one correspondence between the possible values of certain 
quantities'' in two systems. As suggested in articles \cite{ENT,ENT1} that such correspondence can be called observables' entanglement (and it has some features of state entanglement).     

In formal mathematical calculations of sections 2,3 the notations ${\cal S}$ and ${\cal M}$ lose their meaning and  we consider two arbitrary physical systems  $S_1$ and $S_2$ in the states $|\psi\rangle$ and $|\xi\rangle$ belonging to the state spaces ${\cal H}_1$ and ${\cal H}_2.$ Initially the compound system $(S_1,S_2)$ is in the state 
\begin{equation}
\label{VN0}
|\Phi_0\rangle = |\psi\rangle |\xi\rangle
\end{equation}
belonging to the state space ${\cal H}_1 \otimes {\cal H}_2.$
Consider observables for systems $S_1,S_2$  represented by Hermitian operators $A_1, A_2.$ Let $(|\psi_k\rangle)$ and $(|\xi_k\rangle)$ 
be corresponding bases consisting of eignevectors, $A_1|\psi_k\rangle = a_{1k} |\psi_k\rangle, A_2|\xi_k\rangle = 
a_{2k} |\xi_k\rangle,$ i.e., 
\begin{equation}
\label{VN2}
 A_1=  \sum_k a_{1k} |\psi_k\rangle \langle \psi_k|, \;    A_2=  \sum_k a_{2k} |\xi_k\rangle \langle \xi_k|.    
\end{equation}
To simplify considerations, we assume that these operators have non-degenerate spectra.
 
It is proved \cite{VN} (with the explicit mathematical formula) that there exist a unitary operator $U: {\cal H}_1 \otimes {\cal H}_2 \to {\cal H}_1 \otimes {\cal H}_2,$ such that the initial state $|\Phi_0\rangle$ is transferred 
into $A_1A_2$-entangled state:
\begin{equation}
\label{VN1}
 |\Phi \rangle = U |\Phi_0\rangle = \sum_k c_k |\psi_k\rangle |\xi_k\rangle,      
\end{equation}
where 
\begin{equation}
\label{VN1h}
c_k=\langle \psi| \psi_k\rangle
\end{equation}
 (cf. with EPR article \cite{EPR}).
This mathematical result can be reformulated in the language of quantum observables, as following. {\it The outcomes of the observables $A_1$ and $A_2$  are perfectly correlated.} 
The outcome $A_1=a_{1k}$ is perfectly correlated with the outcome $A_2=a_{2k}.$ Thus, measurement of $A_2$ can be considered as indirect measurement of $A_2$ and vice verse.
 
In the light of the above analysis of von Neumann's construction \cite{VN} (chapter 6), the assignment to the systems $S_1$ and $S_2$ the meaning of a system under observation and a measurement apparatus is the purely interpretational issue.  

Now turn to section 3 (chapter 6 \cite{VN}) and set $S_1= {\cal S}, S_2={\cal M}$ and $A_1=A,$ observable on ${\cal S}$ measured 
with ${\cal M}$ and $A_2=M,$ pointer observable. Due to  $AM$-entanglement $M$-outcomes can be identified with the corresponding outcomes 
of the observable $A.$ Although the observer ${\cal O}$ was excluded from calculations, it plays the role: the quantity $M$ is observed by ${\cal O.}$ This mathematical description of  the measurement process matches with RQM Postulate 5:     

\begin{footnotesize}   
Measurement: An interaction between two systems results in a correlation within
the interactions between these two systems and a third one; that is, with respect to
a third system $W,$ the interaction between the two systems $S$ and $F$ is described by a
unitary evolution that potentially entangles the quantum states of $S$ and $F.$
\end{footnotesize}

Subjective nature of observable ${\cal O}$ mentioned (but not so much highlighted) by von Neumann doesn't play any role in the formal 
scheme presented in \cite{VN}.

\section{Conditions of ``probability reproducibility'' vs.  ``to be measured in the same basis''}
\label{PR}

We point out that the measurement processes modeled by von Neumann satisfy the probability reproducibility condition.
We remark that (\ref{VN1h}) implies that 
\begin{equation}
\label{VN1ha}
P(A_{1}= a_{1k}, A_{2}= a_{2k}| \Phi) = |c_k|^2= |\langle \psi| \psi_k\rangle|^2= P(A_{1}= a_{1k}| \psi).
\end{equation}
And by using equalities (\ref{ENT2at}), (\ref{ENT2at1}) we obtain 
\begin{equation}
\label{VN1hb}
P(A_{1}= a_{1k}| \psi) = P(A_{2}= a_{2k}| \Phi),
\end{equation}
We remark that 
$$
P(A_{2}= a_{2k}| \Phi)= |\langle (I \otimes E_{A_{2}})  U \psi \xi|  U \psi \xi\rangle |^2=
|\langle U^\star (I \otimes E_{A_{2}}) U \psi \xi|  \psi \xi\rangle |^2.
$$
Here $U= U(T)= e^{-iT H/\hbar}.$ Set $A_2(T)= U^\star(T) (I \otimes A_{2}) U(T).$ Thus, we obtain the probability reproducibility condition:     
\begin{equation}
\label{VN1hc}
P(A_{1}= a_{1k}| \psi) = P(A_{2}(T)= a_{2k}| \psi \xi),
\end{equation}
or by selection $a_{1k}= a_{2k}= a_{k},$ and setting $A_1= A,A_2=M,$ we get
\begin{equation}
\label{VN1hd}
P(A= x| \psi) = P(M(T)= x| \psi \xi),
\end{equation}
for any $x=a_k.$ 

Thus, if one restricts measurement processes to those described in sections 2, 3 (chapter 6 \cite{VN}), then the 
condition of probability reproducibility in {\bf Postulate 6[OIT]} is  redundant.

Generally measurement process doesn't guarantee probability reproducibility. Such measurements are noisy. Therefore it is not surprising 
that intersubjectivity postulate can be violated. Measurements performed by $O_1$ and $O_2$ are disturbed by noises and their outcomes need not coincide. 

It is important to note that the class of measurement processes satisfying the probability reproducibility condition  doesn't coincide
with von Neumann's measurement processes described in \cite{VN}. To discuss this important issue, we have to refer to the notion of generalized quantum observable, POVM. 

Consider only discrete POVMs valued in finite sets, $X=\{x_1,..,x_m\}.$  POVM is a map $x \to  \Pi(x):$ for $x \in X,  \Pi(x)$ is a positive contractive
self-adjoint operator:  $0\le \Pi(x)\le I,$ called an {\it effect}; the family of  effects form the resolution of unity 
$\sum_x \Pi(x)=I .$
This map defines an operator valued measure on algebra of all subsets of set $X.$
For $O \subset X,$ $\Pi(O)= \sum_{x \in O} \Pi(x).$  POVM  $\Pi$ represents statistics of  measurements:   
\begin{equation}
\label{z1}  
P(\Pi =x  |\psi) =  \langle \psi | \Pi(x) |\psi \rangle . 
\end{equation}
The resolution of unity  condition is the  operator-measure counterpart of the condition normalization by 1 for usual probability measures. Any observable $A$ given by Hermitian operator can also be represented as POVM of the special type -- PVM 
$E_A=(E_A(x)).$

Any measurement process given by quadruple (\ref{MA00}) generates a generalized observable 
given by POVM
\begin{equation}
\label{MA3}
\Pi(x)=  \langle \xi | E_{M(T)}(x)|  \xi \rangle.
\end{equation} 
As was shown by Ozawa \cite{O1}, any POVM, generalized observable, can be generated in this way, from some quantum measurement process.
The tricky point which is often ignored in quantum information theory is that the same observable, POVM, can be generated by a variety of 
measurement processes. 

Now we turn to the probability reproducibity condition. As was shown in \cite{OIT}, a measurement process satisfies to this condition 
if and only if it generates POVM, as (\ref{MA3}), which coincides with PVM of the operator $A,$ i.e., for any $x \in X,$
$\Pi(x)= E_A(x).$ We repeat once again that a plenty of measurement processes can generate the same PVM. 

The violation of the probability reproducibity condition implies that a measurement process generates POVM which is not PVM of the von Neumann observable $A.$    

The original Postulate 6 [RQM] doesn't involve the probability reproducibility condition. Thus, there are the 
following situations. The general theory of  quantum measurement processes allows outcomes to be different, but Postulate 6 \cite{RQMI}
would still require them to be the same. One of the possibilities to resolve this discrepancy is to proceed with the original Postulate 
6 [RQM], but with von Neumann's measurement processes \cite{VN}. Another possibility is to  restrict the class of quantum measurement processes to those satisfying the probability reproducibility condition. 

This is the good place to turn to Ozawa's remark that 
\begin{footnotesize} ``the assumption ``$O_1$ measures $S,$ and $O_2$ also measures $S$ in the same basis'' usually means that both $O_1$'s measurement and $O_2$'s measurement satisfy the probability reproducibility condition for the observables determined by ``the same basis''.''\end{footnotesize} 

My reply to Ozawa's remark was that  \begin{footnotesize} ``Generally you are right, but if we look at the problem more carefully, I think that we should impose the probability reproducibility condition additionally. If we proceed in the quantum measurement framework and operate with measurement processes, then generally we get two generalized observables $M_1$ and $M_2.$ And what does it mean for them, for POVMs, to be measured in the same basis? Only under the probability reproducibility  condition, you proved that in fact these are von Neumann observables given by Hermitian operators. Hence,  one really can speak about measurement in the same basis. But this condition becomes redundant - due to the impose of the probability reproducibility condition.''\end{footnotesize}

This mini-debate highlighted the need of clarification of the condition to ``measurement in the same basis''. And I completely agree with Ozawa's reply:

\begin{footnotesize}
``Postulate 6 ... is ambiguous in what is meant by ``the same basis''.  There are only two possible interpretations:
\begin{itemize}
\item (A)  the same `orthogonal' basis;
\item (B)  the same `not necessarily orthogonal' basis.
\end{itemize}
For interpretation (A), quantum mechanics proves the postulate (the postulate is redundant).
For interpretation (B), quantum mechanics provides a counter example for the postulate (the postulate is not correct).
Note that if the output probabilities for two observers are not the same, it is obvious that they do not measure system $S$ in `the same basis'. 
Thus, the ambiguity is only as above.''  
\end{footnotesize} 
 
We also remark that we considered measurement processes for von Neumann observables, operator $A$ is Hermitian. As is pointed out in 
article \cite{OIT}, OIT does not hold for measurements of generalized observables. In  \cite{OIT} one can find an example of generalized observable $A$ represented by POVM $A=(A(x),$ where $A(x), x \in X,$ are effects,  and two compatible measurement processes $M_1(t)$ and 
$M_2(t)$ satisfying the probability reproducibility condition, but violating the intersubjectivity condition. (Here `compatibility' has the meaning of compatibility of generalized observables). Thus, for generalized observables, one should give up intersubjectivity. As was already mentioned, generalized observables are unsharp and their measurements are noisy. Therefore, one can't expect coincidence of the outcomes of measurements, even for measurement processes satisfying the probability reproducibility condition.   

\section*{Acknowledgments} 

This study was partially supported by  the EU-project: CA21169 - Information, Coding, and Biological Function: the Dynamics of Life (DYNALIFE). I would like to thank Masanao Ozawa for numerous lessons on quantum measurement theory, discussions on intersubjectivity of quantum measurements, and hospitality during my visits to Nagoya during the last 20 years. I also would like to thank Christopher 
A. Fuchs for 25 years of discussions on the QBism's place within quantum theory. (In fact, my interest to intersubjectivity in RQM was ignited by analysis of the intersubjectivity problem in QBism in the view of OIT  \cite{Qozawa}; 
cf. \cite{QBismReply1, QBismReply2}.)

\end{document}